\documentclass[12pt]{article}
\newcommand {\beq} {\begin{equation}}
\newcommand {\eeq} {\end{equation}}
 \newcommand {\ber}{\begin{eqnarray*}}
 \newcommand {\eer} {\end{eqnarray*}}
\newcommand {\bea}{\begin{eqnarray}}
 \newcommand {\eea} {\end{eqnarray}}
\newcommand{\be}{\begin{equation}}
\newcommand{\ee}{\end{equation}}
\newcommand{\eq}[1]{(\ref{#1})}

\def\N {{\cal N}}
\def\ha{\hat{a}}

\newcommand{\Tr}{{\rm Tr}}

\newcommand{\bra}[1]{\langle #1|}
\newcommand{\ket}[1]{|#1 \rangle}
\def\rr{{\rm r}} \def\rs{{\rm s}}\def\ri{{\rm i}}
\def\del{\partial}

\def\a{\alpha}          
\def\b{\beta}           
    
\def\d{\delta}    
\def\e{\epsilon}

\def\l{\lambda} \def\L{\Lambda}
\def\m{\mu}     
     \def\P{\Pi}

\def\th{\theta}

\font\mybb=msbm10 at 12pt

\def\bb#1{\hbox{\mybb#1}}

\newcommand{\sect}[1]{\setcounter{equation}{0}\section{#1}}
\renewcommand{\theequation}{\thesection.\arabic{equation}}

\def\R{{\bb{R}}}
\def\P{{\bb{P}}}

\begin{document}

\hfill{CPHT-RR-013.03.03}

\hfill{LPTENS-03/09}

\hfill{LPTHE-P03-06}

\hfill{NORDITA-2003-15HE}

\vspace{20pt}

\begin{center}

{\LARGE \bf The 3--string vertex and the AdS/CFT duality in the PP--wave limit}
\vspace{30pt}

{\bf Paolo Di Vecchia$^{a}$, Jens Lyng Petersen$^{b}$, 
Michela Petrini$^{c}$, \\ 
Rodolfo Russo$^{d}$,
Alessandro Tanzini$^{e}$
}

\vspace{15pt}
{\small \em
\begin{itemize}
\item[$^a$]
NORDITA,
Blegdamsvej 17, DK-2100 Copenhagen \O, Denmark
\item[$^b$]
The Niels Bohr Institute, Blegdamsvej 17, DK-2100 Copenhagen \O, Denmark
\item[$^c$]
Centre de Physique Th{\'e}orique, Ecole
Polytechnique\footnote{Unit{\'e} mixte du CNRS et de l'EP, UMR
7644}, 91128 Palaiseau Cedex, France
\item[$^d$]
Laboratoire de Physique Th{\'e}orique de
l'Ecole Normale Sup{\'e}rieure,  \\
24 rue Lhomond, {}F-75231 Paris Cedex 05, France
\item[$^e$]
LPTHE, Universit{\'e} de Paris~VI-VII, 4 Place Jussieu 75252 Paris Cedex 05,
France.
\end{itemize}
}

\vskip .1in {\small \sffamily 
divecchia@nbi.dk,
jenslyng@nbi.dk,
petrini@cpht.polytechnique.fr,
rodolfo.russo@lpt.ens.fr, tanzini@lpthe.jussieu.fr}
\vspace{50pt}

{\bf Abstract\\}

\end{center}

We pursue the study of string interactions in the PP-wave background
and show that the proposal of hep-th/0211188 can be extended to a full
supersymmetric vertex. Then we compute some string amplitudes in both
the bosonic and fermionic sector, finding agreement with the field
theory results at leading order in $\l'$.

\setcounter{page}0
\setcounter{footnote}0
\thispagestyle{empty}
\newpage

\sect{Introduction}

The dynamics of type IIB strings on the maximally supersymmetric
PP--wave background~\cite{Blau:2002dy,Blau:2001ne} has been extensively
investigated in the last year. Technically, this has been possible
because in the Green--Schwarz formalism the light--cone string action
contains only free fields~\cite{Metsaev:2001bj,Metsaev:2002re}. 
By using this basic result, various different aspects of string
dynamics on PP--wave backgrounds have been analyzed.

At the conceptual level the motivation for studying strings on
PP--waves is twofold. On one side, it is interesting by itself to have
the possibility to test string theory on a curved background (which is
even not asymptotically flat). On the other side, Berenstein,
Maldacena and Nastase \cite{Berenstein:2002jq} showed that the
dynamics of string theory in these backgrounds is directly connected
to that of (Super) Yang--Mills theories. This relation appears to be a
``corollary'' of the AdS/CFT 
duality~\cite{Maldacena:1997re} and can be used both to derive new
predictions on the gauge theory side and to understand better the
original duality itself. We will focus only on the simplest case of
the maximally supersymmetric PP--wave background. This background can
be obtained~\cite{Blau:2002dy,Berenstein:2002jq} by performing a
Penrose limit~\cite{Penrose,Gueven:2000ru} of $AdS_5 \times S^5$,
where one keeps only
the perturbations with large energy and large spin along a chosen
$SO(2)$ inside the five--sphere. This limit can be
translated~\cite{Berenstein:2002jq} to the ${\cal N}=4$ SYM side by
using the standard AdS/CFT dictionary. The result is a
non--'t-Hooftian limit where $\lambda=g_{YM}^2 N_c \to\infty$. Most of
the gauge invariant operators should decouple in this limit since
their conformal dimensions diverge. On the other hand, a particular
set of operators~\cite{Berenstein:2002jq} (which we call collectively
BMN operators) have a well defined conformal dimension even in the
limit and there is a one--to--one correspondence between this set and the
type IIB string spectrum on the PP--wave
background~\cite{Berenstein:2002jq}. The BMN operators are
characterized by having $J >>1 $, where $J$ is the
R--symmetry charge under the generator singled out in the
Penrose limit on the supergravity side. In formulae the limit is
defined as $\lambda \to\infty$ with $ g_{YM}  \sim const$,
and the surviving operators are characterized by\footnote{The
  relevance of the {\em double} scaling limit has been first stressed
  in \cite{Kristjansen:2002bb,Constable:2002hw}.}
\beq\label{li2}
\Delta, J \to \infty ~~,~{\rm with}~~~~ 
 \lambda'=\frac{\lambda}{J^2} 
\sim const.~,~~  g_2=\frac{J^2}{N_c} \sim const.~,~~
\Delta - J \sim const.
~.
\eeq

In this paper we focus on the study of the $3$--string interaction on
the PP--wave background and discuss the implications of our results for
the PP--wave/CFT duality. A first proposal for the $3$--string vertex
was put forward in~\cite{Spradlin:2002ar}. This vertex was
subsequently corrected and completed in various
papers~\cite{Spradlin:2002rv}, \cite{Pankiewicz:2002gs}
and~\cite{Pankiewicz:2002tg}. The basic idea of these papers is
to apply the techniques used in the flat space case
\cite{Cremmer:1974jq,Cremmer:1974ej,Green:1983tc,Green:1983hw} also to
the PP--wave background. The explicit form of the string interaction is
obtained in two steps. First one rewrites in terms of string
oscillators the delta functional ensuring the smoothness 
of the string world--sheet. This yields a $3$--vertex that is
invariant under all the
continuous kinematical symmetries\footnote{In the light--cone
quantization the symmetries are divided in two types: the symmetries
that preserve the light--cone conditions are called kinematical, and
all the other are dynamical. There is a clear physical reason for this
nomenclature. The kinematical generators commute with the condition
$x^+\sim \tau$, so they act at fixed light--cone time and are not
modified by switching on the interaction. All the other generators can
receive corrections and depend non--trivially on the coupling
constant.}. Then following closely~\cite{Green:1983hw}, this
$3$--string vertex has been supplemented by a prefactor so that it
transforms correctly also under the dynamical symmetries. The final
formulation of the prefactor has been given
in~\cite{Pankiewicz:2002tg}. However, it has been noticed
in~\cite{Chu:2002eu} that this approach does not constrain completely
the form of the $3$--string vertex and that a different starting point
is possible. The basic idea is to use the same techniques as in
\cite{Cremmer:1974jq,Cremmer:1974ej,Green:1983tc,Green:1983hw} but to
build the PP--wave string vertex on the unique vacuum of the theory.
In~\cite{Chu:2002wj} this approach has been used to derive a
$3$--string vertex satisfying all the kinematical constraints.

In Section~2 we show that there is a simple supersymmetric
completion for the vertex of~\cite{Chu:2002wj} 
which seems very promising for two reasons. 
On one side, as discussed in the paragraph~\ref{psusyc}, this
supersymmetric vertex shares some striking similarity with the
behaviour of supergravity on $AdS_5 \times S^5$. 
This suggest that the $3$--state
interaction on the PP--wave should be compared with the results in 
$AdS_5 \times S^5$
rather than with those of flat--space.
On the other side the vertex
presented here represents a simple way to realize the proposal
of~\cite{Constable:2002hw} for comparing  string and field
theory quantities. In Section~3 we discuss some concrete examples.
As in~\cite{Constable:2002hw}, we stay at leading order in $\lambda'$,
but we consider also BMN operators with vector and fermion impurities
(and not only those with bosonic impurities). 
The proposal of~\cite{Constable:2002hw} has been subsequently
criticized~\cite{Beisert:2002bb,Constable:2002vq} and some alternative
prescriptions have been
proposed~\cite{Gross:2002mh,Janik:2002bd,
Gomis:2002wi,Gomis:2003kj,Pearson:2002zs,Chu:2003qd,Georgiou:2003aa,Chu:2003ji}.
In the Conclusions we review the arguments usually presented 
against~\cite{Constable:2002hw} and the various alternative proposals.
In our opinion none of the proposals made so far can be surely
discarded in the present state, not even the one
of~\cite{Constable:2002hw} that provides the simplest possible setup.
The main point we want to stress is
that the PP--wave/CFT duality has to be treated as close as possible
to the usual AdS/CFT. In fact, the physical states of the PP--wave
background are also present in the full $AdS_5 \times S^5$ geometry;
working in the plane wave approximation means just keeping the leading order
in the limits~\eq{li2}\footnote{Also the
subleading correction beyond the PP--wave
limit has been considered in~\cite{Parnachev:2002kk}.}. 
Thus it would be highly desirable
that the results obtained in the PP--wave geometry could be
interpreted on the gauge theory side by means of the usual AdS/CFT
dictionary. We are still far from this goal, mainly because one
crucial ingredient~\cite{Gubser:1998bc,Witten:1998qj} of  AdS/CFT
duality is clearly spelled only in the supergravity limit ({\em i.e.}
the boundary conditions to impose on the string partition function).
However we think that the vertex presented in this paper is a step
forward in this direction.

\sect{The supersymmetric $3$--string vertex}

\subsection{Our conventions for the free string}

The free motion of a IIB superstring in the maximally supersymmetric
PP--wave background can be described by the sum of the following 
2D actions\footnote{We take $\mu >  0$.}
\begin{equation}
S_b = \frac{1}{4 \pi \alpha'} \int d\tau \int_{0}^{2 \pi
|\alpha|} d \sigma \left[ (\partial_{\tau} x)^2 - (\partial_{\sigma}
x)^2 - \mu^2 x^2 \right]~,
\label{bos45}
\end{equation}
\begin{equation}
S_f = \frac{1}{4 \pi \alpha'} \int d\tau \int_{0}^{2 \pi |\alpha|} d
\sigma \left\{i e(\alpha) \left[{\bar{\theta}} \partial_{\tau} \theta +
\theta \partial_{\tau} {\bar{\theta}} + \theta \partial_{\sigma} \theta + 
{\bar{\theta}} \partial_{\sigma} {\bar{\theta}} \right] - 2 \mu 
{\bar{\theta}} \Pi \theta \right\}~.
\label{act567}
\end{equation}
Lorentz indices have been suppressed for sake of simplicity;
the 8 bosons $x^I$ and the 8 fermions $\theta^a$ are always
contracted with a Kronecker $\delta$ except for the mass term in $S_f$,
where $\Pi = \sigma_3 \otimes 1_{4\times 4}$ appears. As usual, we
indicate with $\a'$ the Regge slope, while $\a = \a' p^+$ is the
rescaled light--cone momentum\footnote{As usual in the study of string
interaction in the light--cone, we work with dimensionful
$(\tau,\sigma)$, so $\a$ can be also seen as parameterizing the length
of the string.} and $e(\a) = 1$ if $\a >0$ and $e(\a) = -1$ if $\a
<0$. This form for $S_b+S_f$ follows directly from specializing to
light--cone gauge the covariant action written in
\cite{Metsaev:2001bj,Metsaev:2002re}\footnote{Often in the literature
a redefinition of the spinor coordinates is
adopted ($ \theta \to {\rm e}^{i\pi/4 } \theta$).}.
It is also useful to rewrite the fermionic action in terms of two real
spinors
\begin{equation}
S_f = \frac{1}{4 \pi \alpha'} \int d\tau \int_{0}^{2 \pi |\alpha|} d
\sigma\,\, i \left\{ e(\alpha)
\left[ \theta^1 \partial_{+} \theta^1 + \theta^2
\partial_{-} \theta^2  \right] - 2  \mu
\theta^1 \Pi \theta^2 \right\}~~,~~ \partial_{\pm} =
\partial_{\tau} \pm \partial_{\sigma}~,
\label{act49}
\end{equation}
where $\theta^{1}$ and $\theta^2$ are related to the complex $\theta$
by 
\begin{equation}
\theta = \frac{\theta^1 + i \theta^2}{\sqrt{2}}~~,~~{\bar{\theta}} =
\frac{\theta^1 - i \theta^2}{\sqrt{2}}~.
\label{the54}
\end{equation}
From Eqs.~\eq{bos45} and~\eq{act567} it is
straightforward to derive the mode expansions, the commutation
relations and the expressions for the free symmetry generators. 
Let us recall here some of these ingredients which we will
use in what follows: $\omega_n = \sqrt{n^2 + (\mu\a)^2}$ is the
frequency of the $n^{\rm th}$ string mode and
\begin{equation}
c_n = \frac{1}{\sqrt{1 + \rho_{n}^{2}}}~~,~~ \rho_{n} =
\frac{\omega_n -n}{\mu \alpha}~.
\label{cnrhon}
\end{equation}
At $\tau=0$ the mode expansion for the fermionic fields\footnote{Our
conventions for the bosonic modes are equal to those
of~\cite{Spradlin:2002ar,Pankiewicz:2002gs}.} is
\begin{equation}
\theta = \theta_{0} + \sqrt{2} \sum_{n=1}^{\infty} 
\left[ \theta_{n} \cos \frac{n \sigma}{|\alpha|} +
\theta_{-n} \sin \frac{n \sigma}{|\alpha|} \right]~,
\label{svexpa7}
\end{equation}
where the modes $\theta_n$ can be rewritten in terms of $b_n$'s that
satisfy canonical commutation relations $\{ b_n, b_{m}^{\dagger} \} =
\delta_{nm}$, $ \{ b_n , b_m \} = \{ b_{n}^{\dagger} , b_{m}^{\dagger}
\} =0$
\begin{equation}
\theta_{0}= \frac{\sqrt{\alpha'}}{2 \sqrt{|\alpha|}}
\left[ (1 + e(\alpha) \Pi ) b_0 + 
e(\alpha) (1 - e(\alpha) \Pi) b_{0}^{\dagger} \right]~,
\label{th0sv}
\end{equation}
\begin{equation}
\theta_{n}= \frac{\sqrt{\alpha'}}{\sqrt{2 |\alpha|}} 
\sqrt{\frac{n}{\omega_{(n)}}} 
\left[ P^{-1}_n b_n + e(\alpha) P_n b_{-n}^{\dagger} \right]~,
\label{Pthnsv}
\end{equation}
\begin{equation}
\theta_{-n}= i \frac{\sqrt{\alpha'}}{\sqrt{2 |\alpha|}} 
\sqrt{\frac{n}{\omega_{(n)}}} 
\left[ P^{-1}_n b_{-n} - e(\alpha) P_n b_{n}^{\dagger} \right]~.
\label{Pth-nsv}
\end{equation}
where $P$ is a diagonal matrix in the spinor space
\begin{equation}
\label{Pdef}
P_{n}^{\pm 1} = c_n \sqrt{\frac{\omega_n}{n}} ( 1 \mp \rho_n \Pi) =
\frac{1}{\sqrt{1-\rho^2_{n}}} (1  \mp \rho_{n} \Pi)~.
\end{equation}
The conjugate fermionic momentum is $\lambda_{conj} = - i\lambda = -i 
\frac{e(\alpha)}{2 \pi \alpha'} {\bar{\theta}}$ and its mode expansion
at $\tau=0$ is 
\begin{equation}
\lambda = \frac{1}{2 \pi |\alpha|} \left\{ \lambda_{0} + \sqrt{2} 
\sum_{n=1}^{\infty} 
\left[ \lambda_{n} \cos \frac{n \sigma}{|\alpha|} +
\lambda_{-n} \sin \frac{n \sigma}{|\alpha|} \right] \right\}
\label{svexpa8}~.
\end{equation}
In terms of the $b_n$'s one has
\begin{equation}
\lambda_{0} = \frac{\sqrt{|\alpha|}}{2 \sqrt{\alpha '}} 
\left[ e(\alpha) (1 - e(\alpha) \Pi ) b_0 + 
(1 + e(\alpha) \Pi) b_{0}^{\dagger} \right]~,
\label{la0sv}
\end{equation}
\begin{equation}
\lambda_{n}= \frac{\sqrt{|\alpha|}}{\sqrt{2 \alpha'}} 
\sqrt{\frac{n}{\omega_{(n)}}} 
\left[e(\alpha) P_n b_{-n} +  P^{-1}_n b_n^\dagger \right]~,
\label{Plansv}
\end{equation}
\begin{equation}
\lambda_{-n}= - i \frac{\sqrt{|\alpha|}}{\sqrt{2 \alpha'}} 
\sqrt{\frac{n}{\omega_{(n)}}} 
\left[ P^{-1}_n b_{-n}^\dagger - e(\alpha) P_n b_{n} \right]~.
\label{Pla-nsv}
\end{equation}
The generator of the $x^+ \to x^+ + const.$ transformation is just the
canonical Hamiltonian multiplied by $e(\a)$, since in the light--cone
gauge $x^+$ is identified with $\tau$ ($x^+ = e(\a) \tau$). In the
fermionic sector $H$ reads
\begin{equation}
H_f = e(\a) \int_0^{2\pi \a}\! d\sigma \left(\sum_{i=1}^{2}
\partial_{\tau} \theta^i \lambda^i - L_f \right)= 
\frac{1}{\alpha} \sum_{n= -\infty}^{\infty} 
\frac{\omega_n}{2} (b_{n}^{\dagger} b_n -b_n b_{n}^{\dagger})~.
\label{hami89}
\end{equation}
A similar expression holds also for the bosonic sector, so that the
vacuum energy of the oscillators cancels and one can write 
\begin{equation}
H = \frac{1}{\alpha} \sum_{n= -\infty}^{\infty} 
{\omega_n} (a_{n}^{\dagger} a_n + b_{n}^{\dagger} b_n)~.
\label{hami}
\end{equation}
In the following we will need the explicit expression for the 16
dynamical supercharges
\begin{eqnarray}\label{qminuso}
Q^- &=& e(\alpha)\sqrt{\frac{\a'\mu}{2 |\alpha|}}  \gamma \cdot
  \left[a_0 \left(1+e(\alpha) \Pi\right) +a_0^{\dagger} \left(1-e(\alpha)
      \Pi\right)\right] \lambda_0 + \\
   &+& \frac{1}{\sqrt{|\alpha|}} \sum_{n=1}^{\infty} \sqrt{n}
   \gamma \cdot \left[a_n^{\dagger} P_n b_{-n}
 + e(\alpha) a_n P_n^{-1}  b_n^{\dagger} + i a_{-n}^{\dagger} P_n b_n
 - i e(\alpha) a_{-n} P_n^{-1} b_{-n}^{\dagger} \right]\;,\nonumber
\\ \nonumber
\bar{Q}^-&=& \sqrt{\frac{\mu |\alpha|}{2\a'}} \gamma \cdot \left[a_0
      \left(1-e(\alpha) \Pi\right) +a_0^{\dagger} \left(1+e(\alpha)
      \Pi\right)\right] \theta_0 + \\ 
   &+& \frac{1}{\sqrt{|\alpha|}} \sum_{n=1}^{\infty}
  \sqrt{n} \gamma \cdot \left[a_n^{\dagger} P_n^{-1} b_n
  + e(\alpha) a_n P_n b_{-n}^{\dagger} + i a_{-n}^{\dagger} P_n^{-1}
b_{-n}
  - i e(\alpha) a_{-n} P_n b_{n}^{\dagger} \right]\;.\nonumber
\end{eqnarray}
Instead of working directly with these expressions it is better to
introduce the following linear combinations 
\begin{equation}
Q = \frac{1}{\sqrt{2}} \left(Q^- + \bar{Q}^- \right)~~,~~
\widetilde{Q} = \frac{i}{\sqrt{2}} \left( Q^- - \bar{Q}^- \right)~.
\end{equation}
The reason is that in the $\mu\to 0$ limit $Q$ contains just left
moving oscillators, while $\widetilde{Q}$ depends only on the right
moving ones so that $Q$ and $\widetilde{Q}$ are the direct
generalization of the supercharges usually considered in flat--space
computations. These charges satisfy the algebra\footnote{The relative
sign between $J^{ij}$ and $J^{i'j'}$ has been noticed in
\cite{Kim:2003zw}} 
\begin{eqnarray}\label{sal1}
& \{Q_{\dot{a}},Q_{\dot{b}}\}=
2 \delta_{\dot{a} \dot{b} } (H+T) ~~,~~~
\{\widetilde{Q}_{\dot{a}},\widetilde{Q}_{\dot{b}}\}=
2 \delta_{\dot{a} \dot{b} } (H-T)&
\\ \label{sal2} 
& \{Q_{\dot{a}},\widetilde{Q}_{\dot{b}}\}=
\mu\left[- (\gamma_{ij}\Pi)_{\dot{a}\dot{b}}J^{ij}+
(\gamma_{i'j'}\Pi )_{\dot{a}\dot{b}}J^{i'j'}\right]~.&
\end{eqnarray}
Notice the appearance of $T$ in \eq{sal1}:
\begin{eqnarray}
T & = & i e(\a) \int_0^{2\pi|\a|} \left[\partial_\sigma x(\sigma)\,
p(\sigma) + \partial_\sigma\theta(\sigma)\, \lambda(\sigma)
\right] d\sigma \nonumber \\ \label{Tdef} & = &
\sum_{n=1}^\infty \frac{n}{\a}
\left[(b_n^\dagger b_{-n} + b_{-n}^\dagger b_n) -
i (a_n^\dagger a_{-n} - a_{-n}^\dagger a_n) \right]~.
\end{eqnarray}
At first sight the presence of $T$ looks strange since it seems that
the operator expressions of the supercharges~\eq{qminuso} do not
realize the PP--wave superalgebra~\cite{Blau:2001ne,Metsaev:2001bj}. 
However, if one rewrites
$T$ in terms of the BMN--oscillators $\hat{a}^i$ and
$\hat{b}^a$ (see \eq{bmn-osc} and \eq{fmn-osc} respectively), 
it is easy to see that $T=0$ when the level matching
condition is imposed. Thus the supercharges in~\eq{qminuso} provide a
representation of the superalgebra in the {\em physical} part of the
whole Hilbert space; however for the subsequent manipulation it is
important to remember that the anticommutation rules among the supercharges
contain the operator $T$.

\subsection{The kinematical symmetries}
As already mentioned in the Introduction, the
standard procedure to construct the three-string interaction vertex
consists of two steps. First, one looks for a state $\ket{V}$ 
in the three-string Hilbert space realising the kinematical
symmetries of the PP--wave background. 
The invariance under all the kinematical symmetries translates
into requiring the continuity of the bosonic and fermionic
coordinates, and the conservation of the bosonic and fermionic
momenta.
In \cite{Chu:2002wj}, it was shown that a solution to the above
constraints (see \eq{eq378} and \eq{feq378} in Appendix A) is given by
\be \label{V-pp}
\ket{V}= \delta\left(\sum_{\rr=1}^3 \a_\rr\right) 
\ket{E_a}\, \ket{E_b}\, ,
\ee 
where $\alpha_{\rr}$ is the light-cone momentum of the $r$-th 
string\footnote{We choose $\a_i > 0,~i=1,2$ and $\a_3 < 0$.} and 
$\ket{E_a}$ (resp. $\ket{E_b}$) is the
contribution from the bosonic (resp. fermionic) oscillators. Both
contributions can be represented as an
exponential acting on the three-string vacuum.

The bosonic part is the same as for the vertex in
\cite{Spradlin:2002ar, Spradlin:2002rv, Pankiewicz:2002gs, Pankiewicz:2002tg}
\beq
\label{kinbos}
\ket{E_a} = \exp{\Big\{ \sum_{\rr,\rs=1}^{3} \sum_{m,n \in {\rm\bf Z}} 
a_{m (\rr)}^{\dagger} N^{\rr
\rs}_{mn} a_{n (\rs)}^{\dagger} \Big\} } \ket{v}_{123}~,
\eeq
where $\ket{v}_{123}= \ket{v}_1 \otimes \ket{v}_2 \otimes \ket{v}_3$
is the tensor product of the three vacuum states (see Eq.~\eq{vcc} for
the definition) and the matrix $N^{\rr \rs}_{mn}$
\cite{Spradlin:2002ar} is defined in the Appendix~A.

The fermionic contribution reads 
\bea\label{E-la} 
\ket{E_b} & = & \exp
\left\{ \frac{1 + \Pi}{2}
\left[\sum_{\rr,\rs=1}^3\sum_{m,n=1}^{\infty} b^{\dagger}_{-m(\rr)}
Q_{mn}^{\rr\rs} b^{\dagger}_{n(\rs)} - \sqrt{\a'} \Lambda
\sum_{\rr=1}^3\sum_{m=1}^{\infty} Q^\rr_m b^{\dagger}_{-m(\rr)}
\right] \right.\nonumber \nonumber \\ & & \quad+~\left. \frac{1 -
\Pi}{2} \left[\sum_{\rr,\rs=1}^3 \sum_{m,n=1}^{\infty}
b^{\dagger}_{m(\rr)} Q_{mn}^{\rr\rs} b^{\dagger}_{-n(\rs)} +
\frac{\alpha}{\sqrt{\a'}} \Theta \sum_{\rr=1}^3\sum_{m=1}^{\infty}
Q^\rr_m b^{\dagger}_{m(\rr)} \right] \nonumber \right\} \\ & &
\times~\exp \left\{- \sum_{\ri=1}^2
\sqrt{\frac{\alpha_{\ri}}{|\alpha_3|}} b^{\dagger}_{0(\ri)}
b^{\dagger}_{0(3)} \right\} \ket{v}_{123}~, 
\eea 
where 
\be 
\L \equiv \a_1 \l_{0(2)} - \a_2 \l_{0(1)}, \quad 
\Theta \equiv \frac{1}{\a_3} (\th_{0(1)}- \th_{0(2)}) , 
\quad \a \equiv \a_1\a_2\a_3~.
 \ee 
The matrices $Q$ are diagonal in the spinor space and are defined as
(we use the notations of~\cite{Pankiewicz:2002gs})
\bea 
&&Q^{\rr\rs}_{mn}\equiv e(\a_\rr) \sqrt{\frac{|\a_\rs|}{|\a_\rr|}}\;
    [U_{(\rr)}^{1/2} C^{1/2} N^{\rr\rs} 
C^{-1/2} U_{(\rs)}^{1/2}]_{mn},\label{defQ1} 
\\ && \label{defQ2}~
Q^{\rr}_m \equiv \frac{e(\a_\rr)}{\sqrt{|\a_\rr|}} [U_{(\rr)}^{1/2}
  C_{(\rr)}^{1/2} C^{1/2} N^\rr]_m~.  
\eea 
As vacuum state $\ket{v}$, we choose the state of minimal
(zero) energy, which is annihilated by all annihilation modes
\beq \label{vcc}
a_{n(\rr)} \ket{v}_\rr =0~, \quad b_{n(\rr)} \ket{v}_\rr =0 \quad \forall n.  
\eeq 
This choice of vacuum represents the main difference between the
proposals in \cite{Chu:2002eu,Chu:2002wj} and
\cite{Spradlin:2002ar,Spradlin:2002rv,Pankiewicz:2002gs,Pankiewicz:2002tg}\footnote{In
\cite{Spradlin:2002ar,Spradlin:2002rv, Pankiewicz:2002gs,
Pankiewicz:2002tg} the fermionic part of the vertex is built by acting
on the ground state $|0\rangle$ \beq a_{n(\rr)} \ket{0}_\rr =0~~
\forall n~, \quad b_n |0\rangle =0~~ \forall n \neq 0, \quad \quad
\th_0 |0\rangle =0~. \label{fvacuum} \eeq}. Notice that the last line of~\eq{E-la} is
just a rewriting of the zero--mode structure proposed
in~\cite{Chu:2002eu}. In~\cite{Chu:2002eu} the idea to construct
the vertex on the vacuum~\eq{vcc} was suggested by the study of a
$Z_2$ symmetry present in the supergravity background. The structure
of~\eq{E-la} in the $b_0$ Hilbert space is the origin of  the
differences between the two vertices. This structure can also be
understood in a different way by considering the path integral
treatment of the fermionic ``zero--modes'' $b_0$.  In Eq.~\eq{E-la}
these are treated on the same footing as all the other modes. This is
a very natural approach in the PP--wave background because $b_0$ is
not really a zero--mode, but it is an harmonic oscillator exactly as
the stringy modes (the only difference is that its energy is just
$\mu$ and is independent of $\a'$). On the contrary in
\cite{Spradlin:2002ar,Spradlin:2002rv,Pankiewicz:2002gs,Pankiewicz:2002tg}
the $(\theta_0,\lambda_0)$ sector has been treated in the same way as
in flat space. Thus in their case $\ket{E_b}$ has the same structure
found in the flat space vertex~\cite{Green:1983hw}. In particular it
contains an explicit $\delta$--function ($\sum_\rr \lambda_{0 (\rr)}$)
imposing a selection rule: string amplitudes are zero, unless the
external states soak up the fermions present in the
$\delta$--function. From the path integral point of view this result
is natural only in flat space. In this case, in fact, the $\l_0$'s are
genuine zero--modes appearing only in the measure and can not be
saturated by the weight factor $\exp{(iS)}$.  This yields a
$\delta$--function enforcing a selection rule in the amplitudes. In
the PP--wave case, on the contrary, all modes have a non--vanishing
energy, thus one does not expect any selection rule. The
vertex~\eq{E-la} exactly realizes this expectation, since it enforces
the fermionic constraints with an exponential structure also in the
$b_0$ sector, instead of imposing the constraint as in flat space:
$\delta(\sum_\rr \lambda_{0 (\rr)}) \to \sum_\rr \lambda_{0 (\rr)}
\ket{0}_{123}$.

As a final remark, we would like to mention that an alternative
approach can be used to derive \eq{kinbos} and \eq{E-la}.  We have
verified that this form of the vertex follows uniquely from a path
integral treatment \cite{u}.  We made use of the fact that the system
may be considered as a sum of harmonic oscillators, in particular we
treated the fermionic zero mode oscillators on the same footing as all
others.  Boundary conditions are conveniently specified in a coherent
state basis, equivalent to the Bargmann-Fock formalism
\cite{Faddeev:1975vr,IZ}.  We have also explicitly verified that this
kinematical vertex in fact satisfies the complete set of kinematical
constraints.  In the path integral formalism the prefactor originates
from the fact that the light cone gauge and $\kappa$ gauge conditions
cannot be imposed at the point on the world sheet where the 3 strings
join. Ignoring this complication yields exactly the kinematical part
of the vertex. Rather than attempting to derive the prefactor from
first principles we follow the standard approach of seeking a
supersymmetric completion as described below.

\subsection{Supersymmetric completion}\label{psusyc}

In order for the full supersymmetry algebra to be satisfied at the
interacting level, the kinematical vertex has to be completed with
a polynomial prefactor. An analogous construction also applies
for the dynamical supersymmetry generators $Q$ and $\widetilde{Q}$.
Of course, 
the addition of the prefactor
should not change the properties of the vertex under the kinematical
symmetries. As for the dynamical symmetries, if we define the full
hamiltonian and dynamical charges as $|H_3\rangle$, $|Q_{3\dot{a}}\rangle$ 
and $|\widetilde{Q}_{3\dot{a}}\rangle$~\footnote{For the relation
between these states and the corresponding interaction operators,
see for example Eq.(3.12) of \cite{Spradlin:2002ar}}, 
they have to satisfy
\bea\label{qq}
\sum_{\rr=1}^3  Q_{(\rr) \dot{a}} \ket{Q_{3\dot{b}}} & + &
\sum_{\rr=1}^3  Q_{(\rr) \dot{b}} \ket{Q_{3\dot{a}}} = 
2 \delta_{\dot{a}\dot{b}} \ket{H_3}~, \\
\sum_{\rr=1}^3 \widetilde{Q}_{(\rr) \dot{a}}
\ket{\widetilde{Q}_{3\dot{b}}} & + & 
\sum_{\rr=1}^3 \widetilde{Q}_{(\rr) \dot{b}} \ket{\widetilde{Q}_{3\dot{a}}} =
2 \delta_{\dot{a}\dot{b}} \ket{H_3}~,\\  \label{qq2}
\sum_{\rr=1}^3  Q_{(\rr) \dot{a}} \ket{\widetilde{Q}_{3\dot{b}}} & + &  
\sum_{\rr=1}^3 \widetilde{Q}_{(\rr) \dot{b}} \ket{Q_{3\dot{a}}} =0~.
\eea
To our knowledge there is no way to
derive the prefactor from first principles, the standard approach
being to write a suitable ansatz and then check that it is invariant
under all symmetries~\cite{Green:1983tc,Green:1983hw}.
To proceed further, some physical inputs are then required.
%
In~\cite{Pankiewicz:2003kj} a supersymmetric completion of the
kinematical vertex \eq{V-pp} has been obtained by requiring 
the continuity in the flat space $\mu\to 0$ limit.
This forces to assign an even $Z_2$ parity to the state
$\ket{0}$~\eq{fvacuum}. In this case the string vacuum has to be
$Z_2$--odd because $\ket{v}$ and $\ket{0}$ have opposite
parity~\cite{Chu:2002eu}. 
Correspondingly, the prefactor proposed in~\cite{Pankiewicz:2003kj}
is also $Z_2$--odd, ensuring the $Z_2$--invariance of the interaction
vertex. This vertex has been shown in~\cite{Pankiewicz:2003ap}
to be equivalent to that of~\cite{Spradlin:2002ar,Spradlin:2002rv,Pankiewicz:2002tg}.

Here we present a different approach: following the gauge theory
intuition, the vacuum state of the string Fock space is defined to be
{\it even} under the discrete $Z_2$ symmetry. 
Thus we are led to give up the continuity of
the flat space limit $\mu\to 0$ for the string interaction. Also in
this case it is possible to build a string vertex that is invariant
under the $Z_2$ transformation. In this case, this symmetry is realized 
explicitly, {\em i.e.} both the interaction and the vacuum state are
$Z_2$ invariant at the same time. 
A very simple way to realize the supersymmetry algebra 
is to act on
the kinematical vertex with the free part of the hamiltonian and the 
dynamical charges
\begin{equation} \label{nqh}
\ket{H_3} = \sum_{\rr=1}^3 H_\rr \ket{V}~~,~~ 
\ket{Q_{3\dot{a}}} = \sum_{\rr=1}^3 Q_{r \dot{a}}\; \ket{V}~~,~~
\ket{\widetilde{Q}_{3\dot{a}}} = \sum_{\rr=1}^3 \widetilde{Q}_{\rr
\dot{a}}\; \ket{V}~.
\end{equation}
With this ansatz the relations \eq{qq} and \eq{qq2} are a direct
consequence of the free--theory
algebra~\eq{sal1}--\eq{sal2}\footnote{One 
has to set $T=0$ to get the usual
commutation relations and use the fact that the
kinematical vertex is annihilated by generators $J^{ij}$ and
$J^{i'j'}$, since it is $SO(4) \times SO(4)$ invariant.}.
It only remains to check that the full ansatz still satisfies the
kinematical constraints. To this purpose we need the
explicit expressions for the prefactors in \eq{nqh} and to prove that
the various constituents commute with the constraints in \eq{eq378}, 
\eq{feq378}. 
Commuting 
the annihilation modes in $H$ and $Q$ through the kinematical vertex,
we obtain 
\bea
\ket{H_3} &=& - \frac{\a'}{\alpha}
(1-4\mu \a K) \left[ \frac{1}{4} ({\cal K}^2 + \widetilde{\cal K}^2) 
+ \frac{1+\Pi}{2} \, 
{\cal W}_\Lambda  {\cal Y}_\Lambda
 +  \frac{1- \Pi}{2} \, {\cal W}_\Theta {\cal Y}_\Theta
 \right] \ket{V}~, \nonumber\\ \label{expfo}
\ket{Q_{3\dot{a}}} &=&
-\frac{\a'}{\sqrt{2}\a}
(1-4\mu \a K)\, \,{\cal K}^I \gamma^I_{\dot{a}b}
\left({\cal Y}_\Lambda - {\cal Y}_\Theta \right)_b \ket{V}~, \qquad
\qquad \\ \nonumber
\ket{\widetilde{Q}_{3\dot{a}}} &=& -i\frac{\a'}{\sqrt{2}\a}
(1-4\mu \a K)\, \,\widetilde{\cal K}^I \gamma^I_{\dot{a}b}
\left({\cal Y}_\Lambda + {\cal Y}_\Theta \right)_b \ket{V}~,
\eea
where $K = -(1/4)\, B \Gamma^{-1} B$ (see Appendix~A for the definition of
the matrix $\Gamma$). This remarkably simple form has a  very similar 
structure to what was proposed in the string bit formalism 
\cite{Verlinde:2002ig,Zhou:2002mi,Vaman:2002ka,Pearson:2002zs}.
As in \cite{Spradlin:2002ar}, the contribution of the bosonic
oscillators is contained in 
\begin{equation}
{\cal K} = {\cal K}_0 + {\cal K}_+ +{\cal K}_-\,,\qquad
\widetilde{\cal K} = {\cal K}_0 + {\cal K}_+ -{\cal K}_-~.
\end{equation}
where 
\bea
\label{prefbos}
{\cal K}_0 & = &\P-i\mu\frac{\a}{\a'} \R =
\sqrt{\frac{2}{\a'}}\sqrt{\m\a_1\a_2}
\bigl(\sqrt{\a_1}a_{0}^{\dagger (2)}-\sqrt{\a_2}
a_{0}^{(1) \dagger}\bigr)\,, \\
{\cal K}_+ & = & -
\frac{1}{\sqrt{\a'}} \frac{\a}{1-4\mu\a K}
\sum_{\rr=1}^3\sum_{n=1}^{\infty} 
\left[ \frac{1}{\a_{\rr}} (C C_{(\rr)}^{1/2} U^{-1}_{(\rr)}
  N^{\rr})_n \right]
a^{\dagger}_{n (\rr)}\,, \\
{\cal K}_- &=& - \frac{i}{\sqrt{\a'}} \frac{\a}{1-4\mu\a K}
\sum_{\rr=1}^3\sum_{n=1}^{\infty}
\left[\frac{1}{\a_{\rr}} (C C^{1/2}_{(\rr)} N^{\rr})_n \right]
a^{\dagger}_{-n (\rr)}\,.
\eea
Similarly, for the fermionic oscillators we 
introduced the operators ${\cal Y}_\Lambda$
and ${\cal Y}_\Theta$
\begin{eqnarray}
\label{Y}
{\cal Y}_\Lambda  & = & \frac{1+\Pi}{2}
\left[\Lambda - \frac{\a}{\sqrt{\a'}} 
\sum_{\rr=1}^3
\frac{e(\a_{\rr})}{\sqrt{|\a_{\rr}|}}
\frac{(C^{1/2}_{(\rr)} C^{1/2} U_{(\rr)}^{-1/2} N^{\rr})_n}{1-4\mu\a K} 
b^{\dagger}_{n (\rr)}  \right]~, \\
{\cal Y}_\Theta & = & \nonumber \frac{1-\Pi}{2} \left[
\frac{\a}{\a'} \Theta +  \frac{\a}{\sqrt{\a'}} 
\sum_{\rr=1}^3
\frac{e(\a_{\rr})}{\sqrt{|\a_{\rr}|}}
\frac{(C^{1/2}_{(\rr)} C^{1/2} U_{(\rr)}^{-1/2} N^{\rr})_n}{1-4\mu\a K}
b^{\dagger}_{-n (\rr)} \right]~,
\end{eqnarray}
and
\bea
\label{W}
{\cal W}_\Lambda & = & - \frac{1+\Pi}{2} \left[ \frac{\a}{\sqrt{\a'}}
\sum_{\rr=1}^3 \frac{1}{\sqrt{|\a_{\rr}|^3}} \frac{(C^{1/2}_{(\rr)}
C^{3/2} U_{(\rr)}^{-1/2} N^{\rr})_n}{1-4\mu\a K} b^{\dagger}_{-n
(\rr)} \right]~, \\ {\cal W}_\Theta & = & \nonumber \frac{1-\Pi}{2}
\left[ \frac{\a}{\sqrt{\a'}} \sum_{\rr=1}^3
\frac{1}{\sqrt{|\a_{\rr}|^3}} \frac{(C^{1/2}_{(\rr)} C^{3/2}
U_{(\rr)}^{-1/2} N^{\rr})_n}{1-4\mu\a K} b^{\dagger}_{n (\rr)}
\right]~.  
\eea 
The bosonic constituents of the prefactor, ${\cal K}$ and
$\widetilde{\cal K}$, are the same as in 
\cite{Spradlin:2002rv, Pankiewicz:2002gs,Pankiewicz:2002tg}. 
So we refer to those papers for their commutation
relations with the kinematical constraints.  The fermionic
constituents, \eq{Y} and \eq{W}, on the contrary, are new. Actually,
${\cal Y}_{\Lambda}$ coincides with the ${\cal Y}$ of
\cite{Pankiewicz:2002gs, Pankiewicz:2002tg} in the $\Pi=1$
sector. Since the kinematical constraints are diagonal in the spinor
indices the results in \cite{Pankiewicz:2002gs, Pankiewicz:2002tg}
still hold in our case. In Appendix B, we show that also 
${\cal Y}_{\Theta}$, ${\cal W}_{\Lambda}$ and ${\cal W}_{\Theta}$
commute with the kinematical constraints.


At first sight, it may appear surprising that the simple
proposal~\eq{expfo} is fully consistent, because it is at most
quadratic in the fermionic oscillators. In fact, in flat space it is
crucial to have up to eight fermionic insertions in the prefactor in
order to ensure the reality of the interacting
Hamiltonian~\cite{Green:1982tk}. The point is that, when $\mu=0$ the
real and the imaginary part of $\theta$ always appear quadratically
and never mix. Thus the map $\theta^2 \to - \theta^2$ leaves the
action invariant. The invariance of the flat space vertex under this
change requires that the coefficient of $\Lambda^k$ in the prefactor
is related to the complex conjugate of the coefficient of
$\Lambda^{8-k}$. In the PP--wave case the situation is very different
and already the free action~\eq{act567} is 
not invariant under $\theta^2 \to - \theta^2$, as it is
clear from the formulation given in Eq.~\eq{act49}. In order to restore
this symmetry one has to send at the same time $\Pi$ into $-\Pi$. The
functional form of the vertex presented here is unchanged under the
transformation $(\theta^2,\Pi) \to (-\theta^2,-\Pi)$.
Notice that this difference has an important physical meaning. In flat
space one knows that closed string amplitudes are factorized in two
independent pieces involving left and right moving respectively. The
requirement of having a real vertex according to the
definition~\cite{Green:1982tk} is equivalent to the holomorphic
factorization of physical amplitudes. When $\mu\not=0$ this
factorization breaks down and also the exchange $\theta^2
\to -\theta^2$ stops being a symmetry.


Let us now add some comments on the physical meaning of our
proposal~\eq{expfo}. Actually the simple possibility of
supersymmetrizing the kinematical vertex just by dressing it with the
free form of the dynamical supercharges also exists both in the flat
space case and in the approach
of~\cite{Spradlin:2002ar,Spradlin:2002rv,Pankiewicz:2002gs,Pankiewicz:2002tg}.
However in the flat space case the vertex ansatz~\eq{nqh}
yields trivial {\em on--shell} amplitudes. In fact, in order to derive
the $S$--matrix elements from $\ket{H_3}$ one has first to take care of
the $x^+$ dependence of the vertex. This dependence is not manifest in
the formulae usually written, because it is standard to derive the
string vertex at fixed interaction time $\tau=0$. The general form for
$\ket{H_3}_{\tau_i}$ can be easily derived from $\ket{H_3}_{\tau =0}$
by evolving the oscillators in the vertex with the free
Hamiltonian ($H^{(2)}$). For on--shell external states the usual
recipe~\cite{Cremmer:1974jq,Cremmer:1974ej} is that the $S$--matrix
is derived by integrating over the interaction time
\beq \label{fsa}
A_{phys.} = \int_{-\infty}^\infty  {}_{123}\bra{state}
H_3\rangle_\tau \;d\tau = \delta\left(\sum_{r=1}^3 H^r_{2}\right) 
{}_{123}\bra{state} H_3\rangle_{\tau=0}~.
\eeq
This shows that one has the possibility to redefine the off--shell
vertices by adding terms with the structure of~\eq{nqh}, but these
terms will not contribute to the $S$--matrix.
In the present case one is not interested in S-matrix elements but in matrix
elements of the interaction hamiltonian, and these are picked out by the
short time treatment. In any case the S-matrix energy conservation would
correspond to a similar delta function on conformal dimensions on the field
theory side, which is clearly unreasonable.

Further evidence supporting~\eq{nqh} and the interpretation just
proposed comes from what is known of supergravity on $AdS_5\times
S^5$. The vertex derived here should play the same role as the cubic
{\em bulk} couplings derived from the compactification of IIB theory
on $AdS_5\times S^5$~\cite{Lee:1998bx}. We should then be able to
compare the results of our $\ket{H_3}$ for supergravity states with
the (leading order in $J$) results obtained in $AdS_5\times S^5$. 
It is interesting to notice that the bulk vertices obtained
in~\cite{Lee:1998bx} for $3$ scalars are indeed proportional to
$\Delta_1 + \Delta_2 - \Delta_3$, exactly as the string $3$--point
functions derived from $\ket{H_3}$~\eq{nqh}. It is also important to notice
that in AdS/CFT this factor is necessary to have meaningful $3$--point
functions dual to extremal  correlators ($\Delta_1 +
\Delta_2 = \Delta_3$)~\cite{D'Hoker:1999ea}. Indeed,
in the extremal case the integration over the $AdS_5$ position of the
bulk vertex is divergent~\cite{Freedman:1998tz} and cancels exactly
the factor $\Delta_1 + \Delta_2 - \Delta_3$ in the bulk vertex,
leaving the expected field theory result. This pattern seems to be a
general feature of extremal correlators  and not just a 
peculiarity of correlators among
chiral primaries\footnote{See for instance \cite{Ghezelbash:1998pf},
where the correlator between one scalar and two spinors was
considered.}. 
Thus it is natural to expect that the $3$--point vertices
among supergravity BMN states both in AdS and in the PP--wave
background are proportional to the $\sum_\rr H_2^\rr$.
Notice that the vanishing of the 
energy conserving amplitudes is also one of the requirements of the holographic
string field theory proposed in \cite{Dobashi:2002ar,Yoneya:2003mu}\footnote{
In \cite{Dobashi:2002ar,Yoneya:2003mu} it was also noticed that the fermionic
action can be made $SO(8)$ symmetric by the redefinition 
$(\theta_1,\theta_2) \rightarrow (\theta_1,\Pi \theta_2)$. 
This symmetry is satisfied by our vertex \eq{expfo}.}. 

As a final comment let us come back to the prescription for deriving the
physically meaningful data from $\ket{H_3}$. In the next section, we
will check that the idea~\cite{Constable:2002hw} of comparing field
theory results with the interacting hamiltonian matrix element divided
by $\sum_\rr H_2^\rr$ (see \eq{dual}) works very well at leading
order in $\lambda'$. This proposal is very natural since it is
strongly reminiscent of the formula obtained in quantum mechanics
describing the first order transition probability induced by a
perturbation acting for a very short time. It is then likely that this
prescription has to be corrected in order to check the duality at
higher orders in $\lambda'$. 
It would be interesting to see whether the analogy with quantum
mechanics can be helpful in this generalization, following for
instance the ideas in \cite{Beisert:2002ff}.
One can say that in the PP--wave/CFT
duality we are in the opposite situation with respect to the usual
AdS/CFT computations: we have an explicit and complete expressions for
the $3$--states coupling (while in AdS it is a challenging computation
even to derive some of these couplings), but we do not
have a completely clear prescription to relate them to the field theory side.
One needs
the analogue of the prescription in~\cite{Gubser:1998bc,Witten:1998qj}.
Usually in AdS computations one uses
the bulk--to--boundary propagator, but no simple
analogue of this ingredient has been found in the PP--wave case so far.

\sect{Comparison between string and field theory results}

The PP-wave vertex discussed in the previous section 
is in agreement with the proposal of 
\cite{Constable:2002hw} for comparing the string theory interaction
with the three--point correlators of ${\cal N}=4$ SYM theory.
%
This proposal is motivated by the standard AdS/CFT dictionary between bulk
and boundary correlation functions~\cite{Gubser:1998bc}: since the
light-cone interaction vertex on the PP-wave in \eq{nqh} can be
understood as the generating functional of the correlation functions
among string states, it is natural to put it in correspondence with
the correlation functions of the dual field theory operators. 
A more general motivation for
this proposal was indeed provided in
\cite{Dobashi:2002ar,Yoneya:2003mu} by considering the Penrose limit
of the AdS/CFT bulk--to--boundary formula of
\cite{Gubser:1998bc}.
A specific prescription was proposed in~\cite{Constable:2002hw} for the
leading order in $1/\mu$ 
\be
\label{dual}
\frac{(\langle 1 | \otimes \langle 2| \otimes \langle 3|) ~|H_3
\rangle}{\sum_\rr \left(H_2^{\rr}\right)}
= C_{ijk}~, 
\ee
where $C_{ijk}$ is the coefficient appearing in the
correlator among three BMN operators of R-charge $J_i$
\be
\label{3pt}
\langle  \bar O_i(x_i)  O_j(x_j) O_k(x_k) \rangle
=\frac{C_{ijk}} {(x_{ij})^{\Delta^{(0)}_i+\Delta^{(0)}_j-\Delta^{(0)}_k} 
(x_{ik})^{\Delta^{(0)}_i+\Delta^{(0)}_k-\Delta^{(0)}_j}
(x_{jk})^{\Delta^{(0)}_j+\Delta^{(0)}_k-\Delta^{(0)}_i}
}~,
\ee
and depends on the quantum numbers $J_i$
and on the value of the BMN phase factors.
Notice that in \eq{3pt} we write the canonical  
dimensions $\Delta^{(0)}$ of the BMN operators. In fact 
we will be concerned only with the comparison
at the leading order in ${1/\mu}$, {\it i.e.} at the tree level
in the field theory. 

The proposal \eq{dual} has been questioned in the subsequent
literature \cite{Beisert:2002bb,Constable:2002vq}, and other
conjectures were put forward
\cite{Gross:2002mh,Janik:2002bd,Gomis:2002wi,Chu:2003qd}. The main
argument which should invalidate \eq{dual} is the appearance of a
mixing between single and multi-trace BMN operators at the genus-one
order $g_2=J^2/N_c$ of the graph expansion \cite{Bianchi:2002rw,
Arutyunov:2002rs}.
This mixing affects the tree--level evaluation of the field theory
correlators \eq{3pt}. Thus, if taken into account in the dictionary
\eq{dual}, it would spoil the agreement with the string theory
results.  We notice however that similar issues appear already in the
usual AdS/CFT correspondence for {\it extremal} correlators
\cite{D'Hoker:1999ea} (see also ~\cite{Liu:1999kg}). For these
correlators the mixing is enhanced and should in principle be taken
into account in the comparison with the supergravity calculations.
Actually, as was shown in \cite{D'Hoker:1999ea}, this is not the case,
and one gets agreement between {\it single}--trace extremal
correlators and three-point functions of supergravity on $AdS_5\times
S^5$, without invoking any mixing. 
Moreover, following the
arguments of~\cite{Balasubramanian:2001nh}, the identification between
the number of traces and the number of string states should be valid
as long as the SYM operators are not too ``big''. A simple
quantitative characterization of big operators can be derived by
realizing that overlap between single and double traces is of order
$\sqrt{J J' (J-J')}/N$. If this is not negligible in the planar
limit, then we are dealing with big operators. This shows that, even
if the BMN operators are made of an infinite number of fields, they
are never big since $\sqrt{J J' (J-J')}/N \sim g_2/\sqrt{J}\to 0$.
Thus the usual rules
of AdS/CFT should apply: the single trace operators should correspond
to the elementary string states while the multi-trace operators should
be bound states and so they are not present in the spectrum of the
free string. 

Further support to this picture
comes from the fact that computations with multi--trace operators in
the AdS/CFT correspondence seem to be related to string interactions
which are {\em non--local} on the world
sheet~\cite{Aharony:2001pa,Aharony:2001dp}. Since the leading
correlators in the BMN limit \eq{li2} are extremal correlators (for
the supergravity modes) or non-BPS deformations of these (for the
string modes), it is natural to expect that the same features are
shared by the limiting PP-wave/CFT correspondence and that the local
vertex~\eq{expfo} has to be compared with {\it single}--trace
correlators.
  
In this section we will show that this is indeed the case for
the leading order. We underline that we find agreement  
between string theory and field theory results for all kinds
of BMN operator, including those containing vector\footnote{On the
  field theory side, BMN
operators with scalar impurities were first analysed in 
\cite{Kristjansen:2002bb,Constable:2002hw}, while operators 
with vector impurities were first considered in
\cite{Gursoy:2002yy,Klose:2003tw}.} and fermion impurities. 

For the field theory computations, we adopt the formulation
of ${\cal N}=4$ SYM theory in terms of ${\cal N}=2$ multiplets.
This formulation has the advantage to realize explicitly the
$R$-symmetry subgroup $SU(2)_V\times SU(2)_H\times U(1)_J\subset SU(4)$,
where $SU(2)_V, SU(2)_H$ are respectively the internal symmetry groups
of the ${\cal N}=2$ vector multiplet and hypermultiplet.
In this way we can naturally identify the charge under $U(1)_J$ with 
the $J$-charge of the BMN operators, and the $SU(2)_V\times SU(2)_H$ group
with the $SO(4)$ rotations acting on the scalar impurities.
The $Z$ field of the BMN operators is the complex scalar of the 
vector multiplet, while the scalar impurities are given by the four real scalars 
$\phi^{i'}, i'=1,\ldots,4$ of the hypermultiplet. 
The fermionic excitations are associated with the Weyl fermions
$\lambda_{\a}^{u}$ of the vector multiplet and $\bar\psi_{\dot\a}^{\dot u}$
of the hypermultiplet. These fields have in fact both charge $J=1/2$. 
They transform in the fundamental representation $u=1,2$ of $SU(2)_V$
and $\dot u =1,2$ of $SU(2)_H$ respectively.
The quadratic part of the $\N=4$ SYM lagrangian 
is\footnote{We adopt the following convention for the sigma matrices: 
$\sigma^m=(i\sigma^c,{\bf 1})$ and $\bar{\sigma}^m=(-i\sigma^c,{\bf 1})$,
where $\sigma^c$ are the Pauli matrices.}
\bea
\label{act}
{\cal L} &=& \frac{1}{g_{YM}^2}\Tr\Big[\frac{1}{4}F^{mn}F_{mn}
+ D_m\bar Z D^m Z 
+\frac{1}{2}D_m\phi^{i'}D^m\phi^{i'}\\ \nonumber
&&\quad\quad\quad~ + \bar\lambda^u D_m\bar\sigma^m\lambda_u 
+ \bar\psi_{\dot u}D_m\bar\sigma^m\psi^{\dot u} \Big] \ ,
\nonumber
\eea
where the covariant derivative is $D_m\equiv \del_m + [A_m,\cdot]$.
The Green function for the complex field $Z$ is then 
$G(x-x^\prime)={g^2_{YM}/ 4\pi^2(x-x^\prime)^2}$.
The BMN operator associated to the string theory vacuum is
\be
 O_{vac}^{J}(x) = \frac{1}{\sqrt{JN^J}} \Tr\Big[Z^J\Big](x) \ \ ,
\label{ovac}
\ee
with $N=g_{YM}^2 N_c/4\pi^2$. For single impurities we define
\bea
&
O_{i}^{J}(x) = {\N_1}
\Tr\Big[D_i Z Z^J\Big](x),~~~~
O_{i'}^{J}(x) =  \N_1  \Tr\Big[\phi^{i'}Z^J\Big](x), &
\label{o1vec}\\
&
 O_{\a}^J =  \N_1 \Tr\Big[\lambda_\a Z^J\Big](x),~~~~
 O_{\dot\a}^J = \N_1 \Tr\Big[\bar\psi^{\dot\a} Z^J\Big](x), &
\label{o1fer}
\eea
with $\N_1 = 1/\sqrt{2 N^{J+1}}$. The first operator in \eq{o1fer}
corresponds to the insertion of a string fermionic oscillator of
chirality $\Pi = 1$, while the second to the chirality $\Pi=-1$. 
For double impurities we define (in the dilute gas approximation)
\bea\label{o2vec}
&
O_{ji, m}^{J}(x) = \N_2  \sum\limits_{l=0}^{J} q^l\;
\Tr\Big[D_j Z Z^l D_i Z Z^{J-l}\Big](x), & \\
& O_{\a \b, m}^{J}(x) = \N_2 
\sum\limits_{l=0}^{J} q^l\; 
\Tr\Big[\lambda_{\a} Z^l \lambda_{\b} Z^{J-l}\Big](x), & 
\label{o2fer}
\eea
with $\N_2  = (1/2)\,\sqrt{(J+1)N^{J+2}}$.
To simplify  notations, in the above formulae the BMN phase factor is
$q^l=\exp\left(2 \pi i m\frac{l+1}{J+2}\right)$, and
the $R$--symmetry indices of the fermions are suppressed.
The barred operators are defined according to the rules suggested by the
radial quantization \cite{Fubini:1973mf}.
For example, for double--vector
impurities we define\footnote{M. Petrini, R. Russo and A. Tanzini are
  happy to acknowledge collaboration with C.S. Chu and V.V. Khoze about 
the definition of the barred operators in field theory. See 
also~\cite{Chu:2003ji}.}
\be
\label{baro-ij}
\bar O_{ij,m}^J (x) \equiv \N_2 \ 
(r^2)^{J+2}
\sum_{l=0}^{J}  
q^l \Tr \left[
(C_{ik} \bar{D}_{k} \ r^2\bar{Z} ) \bar{Z}^l 
( C_{jh}\bar{D}_{h}\ r^2\bar{Z} )\bar{Z}^{J-l} 
\right] (x) \ ,
\ee
where $C_{ik}(x) =\delta_{ik}-2 x_{i}x_{k}/r^2$
is the tensor associated to the conformal
inversion transformation $x'_{i} = x_{i} / r^2$, with
$\partial x'_{i}  / \partial x_{k} = C_{ik} (x) / r^2 $.
For fermionic impurities we define
\bea
&&\bar{O}_{\a}^{J}(x) \equiv {\N_1} (r^2)^{J+3/2}
\Tr\Big[(\bar{\lambda}\!\not\!\bar x)^\a \bar{Z}^J)\Big](x) \ ,
\label{baro-la} \\
&&\bar{O}_{\dot\a}^{J}(x) \equiv {\N_1} (r^2)^{J+3/2}
\Tr\Big[(\psi\!\not\!x)_{\dot\a} \bar{Z}^J)\Big](x) \ ,
\label{baro-psi}
\eea  
with
$\not \!\bar x\equiv\bar\sigma_k^{\dot\a\a}x^k/r$,
$\not \! x\equiv\sigma^k_{\a\dot\a}x_k/r$ 
and $r\equiv |x|$.

The check of \eq{dual} for BMN operators containing only scalar impurities
has been discussed in detail in
\cite{Kiem:2002xn,Huang:2002wf,Chu:2002pd}. 
We will focus on vector and spinor impurities.
Inspired by the radial quantization analysis, we will evaluate the three-point
correlator \eq{3pt} in the $x_i\rightarrow\infty$ limit. 
In this limit, the following identities hold for vector insertions
\bea\label{dzc}
&&{\N_1^2}\lim_{r\rightarrow \infty}
\langle0|(r^2)^{J+1} \Tr \left[ C_{ik}{\del}^k (r^2 \bar{Z})
\bar{Z}^{J} \right](x)  \Tr \left[ \del'_j Z Z^{J} \right](x')
|0\rangle = \nonumber \\
&& =  \lim_{r\rightarrow \infty} C_{ik} 
\left(\delta_{kj} - \frac{2 x_k x_j}{r^2}\right) = \delta_{ij}
\eea
and
\bea
& &\lim_{r\rightarrow \infty}
\langle0| (r^2)^{J+1} \Tr
\left[C^{ik}{\del}_k (r^2 \bar{Z}) \bar{Z}^{J} \right](x)
 \Tr \left[Z^{J} \right] (x')|0\rangle  \nonumber \\
&& =  \lim_{r\rightarrow \infty}
\del_{k}\left(\frac{r^2}{(x-x')^2}\right) = 0 \ \ .
\label{dzzc}
\eea
From \eq{dzc} and \eq{dzzc} it immediately follows that the vector impurities 
in the BMN operators
behave exactly as the scalar impurities. We are then left
with the computation of the same combinatorial factors studied in 
\cite{Constable:2002hw}. Notice that \eq{dzc} and \eq{dzzc} are consistent
with the string state/operator correspondence.

The fermionic insertions display an analogous behaviour in 
the $r\rightarrow\infty$
limit 
\bea
&&{\N_1^2}\lim_{r\rightarrow \infty}
\langle0|(r^2)^{J+3/2} 
\Tr\Big[(\bar{\lambda}\!\not\!\bar x)^\a \bar{Z}^J)\Big](x)
\Tr\Big[\lambda_\b Z^J\Big]|0\rangle=  \delta^\a_{~\b} \ ,
\label{lac}\\
&&\N_1^2\lim_{r\rightarrow \infty}
\langle0|(r^2)^{J+3/2} 
\Tr\Big[(\psi\! \not\! x)_{\dot\a} \bar{Z}^J)\Big](x)
\Tr\Big[\bar\psi^{\dot\b} Z^J\Big]|0\rangle=  \delta_{\dot\a}^{~\dot\b} \ \ .
\label{psic}
\eea   
This implies that also the tree-level evaluation
of BMN correlators with fermion impurities can be reduced 
to the scalar impurity case, apart from some (important) signs due
to the anticommuting nature of $\lambda$ and $\bar\psi$. 

We now proceed to the comparison of some specific field theory correlators
with the results derived from
the three-string vertex \eq{nqh}. Let us start by considering BMN operators
with one vector and one scalar impurity. From the results in the
literature\footnote{See for instance Eqs.(3.15) and (3.25) in 
\cite{Beisert:2002bb}. At the level of 
planar free field theory there is no difference between the 
single/double--trace correlators in \cite{Beisert:2002bb} and the
single--trace $3$-point functions considered here.}
 and using \eq{dzc} and \eq{dzzc} one can find
\be
\langle \bar O^{J_3}_{i i',n} O^{J_2}_{ii',m} O^{J_1}_{vac} \rangle
\equiv
\lim_{x_3\rightarrow \infty}
\langle \bar O^{J_3}_{i i',n}(x_3) O^{J_2}_{ii',m}(x_2) 
O^{J_1}_{vac}(x_1) \rangle
= \frac{J_2}{N}\sqrt{\frac{J_1J_2}{J_3}}\frac{\sin^2(\pi n y)}{\pi^2(m-ny)^2} \ \ ,
\label{tri-2}
\ee 
where $y\equiv J_2/J_3$.
For operators containing two vector impurities we can also consider the correlator
\be
\langle \bar O^{J_3}_{i i,n} O^{J_2}_{ii,m} O^{J_1}_{vac} \rangle
= \langle \bar O^{J_3}_{i j,n} O^{J_2}_{ij,m} O^{J_1}_{vac}\rangle
+ \langle \bar O^{J_3}_{i j,-n} O^{J_2}_{ij,m} O^{J_1}_{vac}\rangle
\label{sing}
\ee
The last term in \eq{sing} comes from the fact that impurities
with the same vector index can be contracted in two different ways, and
the exchange of two impurities induces a change of sign in the BMN phase
according to the definition \eq{o2vec}.

Let us now see how these results are reproduced from the 
$\mu\rightarrow\infty$ limit of string theory.
Due to the form \eq{nqh} of the prefactor, the non--trivial part
of the proposal \eq{dual} is simply given  by the coefficient $C_{ijk}$.
This justifies the approach of
\cite{Kiem:2002xn,Huang:2002wf,Chu:2002pd}, and  we 
can focus only on the kinematical part
\eq{V-pp} of the $3$--string vertex and rewrite \eq{dual} as 
\be
\langle i | \langle j | \langle k 
|V \rangle = \frac{C_{ijk}}{C^{(0)}_{ijk}} ~,
\label{come}
\ee
where $C^{(0)}_{ijk}=\sqrt{J_1J_2J_3}/N$ is 
the combinatorial factor of the Green function among
three vacuum operators \eq{ovac}. This factor ensures
the same normalization for the two sides of
\eq{come}, since the string overlap ${}_{123}\langle v |V\rangle$
is set equal to one.
For the comparison with field theory it is useful to 
recall the dictionary between the BMN oscillators $\ha$
and those used in the previous section
\beq
\hat a_n = \frac{1}{\sqrt{2}}(a_n - i a_{-n}) \quad ,\quad \quad 
\hat a_{-n} = \frac{1}{\sqrt{2}}(a_n + i a_{-n})~ .
\label{bmn-osc}
\eeq
According to the string state/operator mapping, the relevant
amplitude is
\bea
A^{IJ}_b & = &
\bra{\a_1,\a_2,\a_3}\ha_{n(3)}^I\ha_{-n(3)}^{J}
\ha_{m(2)}^{I}\ha_{-m(2)}^{J} \ket{V}
\label{aijb} \\  \nonumber
& = & \frac{1}{4}\;\bra{\a_1,\a_2,\a_3}
\left(a_{n(3)}^{I}a_{n(3)}^{J} + a_{-n(3)}^{I}a_{-n(3)}^{J}
+i a_{n(3)}^{I}a_{-n(3)}^{J} -i a_{-n(3)}^{I}a_{n(3)}^{J}\right)
 \\  \nonumber && \times~
\left(a_{m(2)}^{I}a_{m(2)}^{J} + a_{-m(2)}^{I}a_{-m(2)}^{J}
+i a_{m(2)}^{I}a_{-m(2)}^{J} -i a_{-m(2)}^{I}a_{m(2)}^{J}\right)
\ket{V} ~.
\eea
We will compute the string theory amplitudes in the $\mu\to\infty$
limit by using the relation $N^{\rr\rs}_{-m-n} = - (U_{(\rr)}
N^{\rr\rs}U_{(\rs)})_{mn}$ and the expansions
\bea
N^{32}_{nm}&\sim& \frac{2}{\pi}\frac{ny^{3/2}\sin(\pi
  ny)}{m^2-n^2y^2}~,~~~{\rm with}~~n,m>0
\label{nlim}
\\
U_{(i)}&\sim& \frac{n}{2\mu\a_i}~, \quad\quad 
U_{(3)}\sim -\frac {2\mu\a_3}{n}~.
\label{ulim}
\eea
Let us first analyze the case $I=i$, $J=i'$. The amplitude~\eq{aijb} then
becomes
\beq\label{323}
A^{ii'}_b = \frac{1}{4} \left[
N^{32}_{nm} - N^{32}_{-n-m}\right]^2 \sim 
y \frac{\sin^2 (\pi n y)}{\pi^2 (m-ny)^2}
~,
\eeq
where in the last step we kept only the leading term in the $\mu\to\infty$
limit. By using \eq{tri-2} this provides a first check of \eq{come}.
If we instead take $I=J=i$,  \eq{aijb} becomes
\bea
A^{ii}_b&=& \frac{1}{2}\left[(N^{32}_{nm})^2 + (N^{32}_{-n-m})^2
+ \frac{1}{2} (N^{33}_{nn} + N^{33}_{-n-n})(N^{22}_{mm} + N^{22}_{-m-m})
\right]\nonumber\\
&\sim& \frac{y}{\pi^2} \sin^2(\pi ny)
\left[\frac{1}{(m-ny)^2}+\frac{1}{(m+ny)^2}\right] ~,
\label{aiib}
\eea
where the two terms in the last parenthesis reproduce
the r.h.s. of \eq{sing}. Notice that the terms proportional to 
$N^{33}$ and $N^{22}$ do not contribute at the leading order.

We now pass to the analogous correlation functions for BMN operators 
with fermionic impurities. By using \eq{lac} and \eq{psic}
their evaluation is reduced to the same combinatorics as for
the scalar impurities case. Thus we have
\be
\langle \bar O^{J_3}_{\a \dot\b,n} O^{J_2}_{\a\dot\b,m} O^{J_1} \rangle
= - \frac{J_2}{N}\sqrt{\frac{J_1J_2}{J_3}}\frac{\sin^2(\pi n
  y)}{\pi^2(m-ny)^2} \ \ , 
\label{fri-2}
\ee
which coincides with \eq{tri-2}.
For operators containing spinors of the same flavour we can also
consider
\be
\langle \bar O^{J_3}_{\a \a,n} O^{J_2}_{\a\a,m} O^{J_1}_{vac} \rangle
= \langle \bar O^{J_3}_{\a\b,n} O^{J_2}_{\a\b,m} O^{J_1}_{vac}\rangle
- \langle \bar O^{J_3}_{\a\b,-n} O^{J_2}_{\a\b,m} O^{J_1}_{vac}\rangle ~.
\label{fing}
\ee
Notice that due to the fermionic nature of the impurities one gets
 a relative minus sign in \eq{fing} with respect to \eq{sing}.

The string amplitudes related to these correlators can be easily
computed from \eq{E-la} by using the dictionary
\beq
\hat b_n = \frac{1}{\sqrt{2}}(b_n + b_{-n}) \quad ,\quad \quad 
\hat b_{-n} = - \frac{i}{\sqrt{2}}(b_n - b_{-n})~ .
\label{fmn-osc}
\eeq
In this case the relevant amplitude is 
\be
A^{ab}_f = 
\bra{\a_1,\a_2,\a_3}\hat b_{n(3)}^{a}\hat b_{-n(3)}^{b}
\hat b_{m(2)}^{a}\hat b_{-m(2)}^{b} \ket{V}
\label{aijf}
 =  -\frac{1}{4}\left[(Q^{23}_{mn})^2+ (Q^{32}_{nm})^2
-2 Q^{23}_{mn}Q^{32}_{nm}\right] ~.
\ee
By using now \eq{defQ1} we get at the leading order in $1/\mu$
\be
Q^{32}_{nm}\sim - N^{32}_{nm} ~, \quad\quad
Q^{23}_{mn}\sim - N^{23}_{-m-n} ~.
\label{QN}
\ee
This shows that in the large $\mu$ limit \eq{aijf} takes the same form
as \eq{323}, in agreement with the field theory result.
Also the correlator \eq{fing} is reproduced by the corresponding
string amplitude
\be
A^{aa}_f = 
\bra{\a_1,\a_2,\a_3}\hat b_{n(3)}^{a}\hat b_{-n(3)}^{a}
\hat b_{m(2)}^{a}\hat b_{-m(2)}^{a} \ket{V}
\label{aiif}
 = \left[-Q^{33}_{nn} Q^{22}_{mm}
+Q^{23}_{mn}Q^{32}_{nm}\right] \sim N^{23}_{-m-n}N^{32}_{nm}.
\ee
This is again in agreement with the field theory result
since
\be
 N^{23}_{mn}N^{32}_{nm}\sim
- \frac{y}{\pi^2} \sin^2(\pi ny)
\left[\frac{1}{(m-ny)^2}-\frac{1}{(m+ny)^2}\right] ~.
\ee
Notice that the agreement in the fermion sector crucially depends 
on the form of our vertex \eq{E-la}. In fact, some fermionic amplitudes
were derived also by using the proposal of~\cite{Pankiewicz:2002tg}
(see the last section of that paper).
These results are quite different from ours because of the different zero
mode structure of their vertex and have not been related to any field theory
result.   

\sect{Conclusions}

In this paper we presented a supersymmetric string vertex that
completes the construction of~\cite{Chu:2002eu,Chu:2002wj}. We argued
that this vertex has all the necessary properties to describe the
local interaction of three strings in the maximally supersymmetric
PP--wave background. We also observe that our vertex might be
seen as an explicit realization of "holographic string field theory"
of \cite{Dobashi:2002ar,Yoneya:2003mu}.
A different supersymmetric completion of the construction of~\cite{Chu:2002eu,Chu:2002wj}
was provided in~\cite{Pankiewicz:2003kj} by requiring the continuity 
of the flat space limit $\mu\to 0$. This vertex has been shown in~\cite{Pankiewicz:2003ap} 
to be equivalent to that in~\cite{Spradlin:2002ar,Spradlin:2002rv,Pankiewicz:2002gs,Pankiewicz:2002tg}.

Let us summarize here the main features of 
our proposal, since it is rather different from the one
of~\cite{Spradlin:2002ar,Spradlin:2002rv,Pankiewicz:2002gs,Pankiewicz:2002tg}.
First, from the very beginning of our construction we gave up the idea
of smoothly connecting our vertex to the one of flat
space~\cite{Green:1983hw} when the parameter controlling the curvature
of the background is sent to zero ($\mu\to 0$). 
One can argue that this limit is singular due to the enhancement
of the isometry group that takes place exactly at the point $\mu=0$.
One consequence of this fact is that for all values of $\mu$
the modes $a_0$ and $b_0$ of the string expansion are  
harmonic oscillators and only for $\mu=0$ one has to
deal with genuine zero--modes $p_0$ and $\lambda_0$.  
For this reason we
treated the modes $a_0$ and $b_0$ on the same footing as all the other
modes in the string expansion.
Moreover, since we do not expect
any $\delta$--function on the energies in the matrix elements, there
is no reason why the
string $2$--point functions should be diagonal at all (perturbative)
orders. It is an open possibility that the identification between
string states and 
single--trace 
field theory operators originally proposed
in~\cite{Berenstein:2002jq} is actually valid also at higher
orders. We took this point of view, since it allows to keep the
physical intution of {\em identifying} the closed string with the
``long'' BMN trace made out of $Z$'s and the string excitations with
the BMN impurities.

In the second part of the paper we used the proposal
of~\cite{Constable:2002hw} and compared the matrix elements of our
$3$--string vertex with the corresponding field theory correlators at
leading order in $\lambda'$. The two results match in all cases. Our
analysis concerns $3$--point correlators of BMN operators with double
insertions of scalar, vector and fermion impurities and provide a test
for the various building blocks of our string vertex in the $\mu\to\infty$
limit. The
proposal of~\cite{Constable:2002hw} has been criticized in the
subsequent literature also from the field theory point of view. It was
argued that the presence of mixing between single--trace and
multi--trace BMN operators had to be taken into account also at the
level of planar computations. This affects the tree--level expression
of the $3$--point correlators and thus spoils the agreement with
string theory results. However, the inclusion of this mixing in the
comparison with bulk amplitudes seems a rather unnatural procedure
from the point of view of the AdS/CFT correspondence. In fact, as we
discussed in Section~3, a similar situation appears also in the
standard AdS/CFT duality in the case of extremal correlators. There
the results found in $AdS_5\times S^5$ supergravity agree with
field theory computations without invoking the mixing.

Let us finally comment on the other methods for comparing field and
string theory that have been
proposed~\cite{Gross:2002mh,Janik:2002bd,Gomis:2002wi,Chu:2003qd}.
There are basically two distinct approaches. In the first one, the
idea is to interpret the relation $\Delta - J = H$ as an operator
equation and to identify the matrix elements of the string Hamiltonian
with those of $\Delta - J$ between multi--trace
operators. Technically this requires a modification of the dictionary
between string states and field theory operators. However this new map
contains several free parameters that are fixed by {\em imposing} the
matching with the string theory results. In the more recent
literature, this has been done by using the string vertex proposed
by~\cite{Spradlin:2002ar,Spradlin:2002rv,Pankiewicz:2002gs,Pankiewicz:2002tg}.
However, it is possible to choose
a different basis in field theory in order to get agreement with the results
of the string vertex \eq{nqh} presented here\footnote{See for example
the first version of \cite{Gross:2002mh}.}. 
In order to get a non--trivial check in this approach one is
obliged to go to higher orders in the parameter controlling the genus
expansion. At this level the computations are rather
involved. Moreover, at subleading order only the field theory answer is
known and is compared with results extrapolated from the tree--level
string vertex by using the quantum mechanical perturbation
theory. Genuine computations on the torus for the mass of string
states remain a challenging open problem. Another approach advocated
in~\cite{Chu:2003qd} is to fix a new dictionary by taking, on the
field theory side, the operators with definite conformal
properties. The comparison with string theory has been done by using
the kinematical vertex of~\cite{Chu:2002wj} supplemented by a
prefactor phenomenologically derived by the inputs coming from
field theory. This prefactor is different from the one presented
here, because it treats the $a_n$ modes with positive and negative
frequency in a different way. Actually
in string computations, like for
instance that of the action of $Q$ on the kinematical
vertex~\eq{expfo}, 
both kinds of modes are treated on the same footing and appear 
always together in the combinations ${\cal
K}$ and $\widetilde{\cal K}$.

The advantage of the setup presented here, beyond its semplicity, is
that there are no ``free parameters'' that can be modified and each
computation represents a test for the proposal. It would be very
interesting to see whether the agreement is preserved when operators
with more than two impurities are considered. The main limitation of
our approach is that it only considers the leading order in
$\lambda'$. Generalizing it beyond this approximation is an important
open problem. On the other hand, since there are two different
proposals for the $3$--string vertex both satisfying the same
super--algebra, it is also important to analyze again the derivation
of $\ket{H_3}$. Clearly to understand the origin of the differences it
is necessary to use in the derivation a dynamical principle (like the
path integral) instead of symmetry arguments. Work is in progress
along these directions. 

\section*{Acknowledgments}

We would like to thank C. Bachas, C.S. Chu and C. Kristjansen 
for interesting discussions and suggestions. 
This work is
supported in part by EU RTN contracts HPRN-CT-2000-00122 and
HPRN-CT-2000-00131. M.P., R.R. and A.T. are supported by European Commission
Marie Curie Postdoctoral Fellowships.

\section*{Appendix A: Definitions}
\renewcommand{\theequation}{A.\arabic{equation}}
\setcounter{equation}{0}

In this Appendix we recall some of the definitions we use in the
paper (we refer to~\cite{Pankiewicz:2002gs} for a more detailed list
of formulae and identities). Let us start with  the mode expansion of the
kinematical contraints
\beq
x_{m(3)} = - \sum_{n = -\infty}^{\infty} \sum_{\ri=1}^{2}
\frac{\alpha_\ri}{\alpha_3 }X^{(\ri)}_{mn} 
x_{n(\ri)} ~~,~~
p_{m(3)}  =  -  
\sum_{n = -\infty}^{\infty} \sum_{\ri=1}^{2} X^{(\ri)}_{mn} p_{n(\ri)}~,
\label{eq378}
\eeq
\beq
\label{feq378}
\theta_{m(3)} = - \sum_{n =  -\infty}^{\infty} \sum_{\ri=1}^{2}
\frac{\alpha_\ri}{\alpha_3 }X^{(\ri)}_{mn} \theta_{n(\ri)}~~,~~
\lambda_{m(3)}  =  -  
\sum_{n = -\infty}^{\infty} \sum_{\ri=1}^{2} X^{(\ri)}_{mn} \lambda_{n(\ri)}~,
\eeq
where the  matrices $X$ are 
\begin{equation}
X^{(\rr)}_{mn}\equiv
\cases{
(C^{1/2} A^\rr C^{-1/2})_{mn}\,, & $m>0\,,n>0$\,, \cr
\frac{\a_3}{\a_\rr}(C^{-1/2}A^\rr C^{1/2})_{-m,-n}\,,&  $m<0\,, n<0$\,, \cr
- \frac{1}{\sqrt{2}}\e^{\rr \rs}\a_\rs (C^{1/2}B)_m \,, & $m>0\,, n=0$\,, \cr
1\,, & $m=0=n $\,,\cr
0\,,& \mbox{otherwise}\, .}
\end{equation}

All the quantities in the equations below are defined for $m,n>0$ 
\begin{eqnarray}
&& C_{mn} = m\d_{mn}\,,\nonumber\\
&& A^{(1)}_{mn} = (-1)^n\frac{2\sqrt{mn}\b}{\pi}\frac{\sin
m\pi\b}{m^2\b^2-n^2}, \nonumber\\
&& A^{(2)}_{mn} =  \frac{2\sqrt{mn}(\b+1)}{\pi}\frac{\sin
m\pi\b}{m^2(\b+1)^2-n^2}, \nonumber\\
&& A^{(3)}_{mn} =  \d_{mn}, \nonumber\\
&& B_m=-\frac{2}{\pi}\frac{\a_3}{\a_1\a_2}m^{-3/2}\sin m\pi\b.
\end{eqnarray}

The matrices $N^{\rr \rs}_{mn}$ and $N^{\rr}_{m}$ 
in the definition of the $Q$'s are given by
\bea
&& N^{\rr\rs}_{mn} =  \delta^{\rr\rs}\delta_{mn}
-2 \left(C_{(\rr)}^{1/2}C^{-1/2} A^{(\rr)\,T}\Gamma^{-1}A^{(\rs)}C^{-1/2}
C_{(\rs)}^{1/2}\right)_{mn}\,,\\
&& N^{\rr\rs}_{-m-n} = - (U_{(\rr)} N^{\rr\rs}U_{(\rs)})_{mn}
\nonumber
~~,\quad 
N^\rr_m = -(C^{-1/2}A^{(\rr)\,T}\Gamma^{-1}B)_m~ ,
\eea
where $C_{(\rs)}= \delta_{mn}\, \omega_{(\rr) m} = \delta_{mn} \sqrt{m^2
+  (\mu \alpha_{(\rr)})}$, $\Gamma
=\sum_{\rr=1}^3 A^{(\rr)}U_{(\rr)}A^{(\rr)\,T}$ and finally 
$U_{(\rr)}= C^{-1}\left(C_{(\rr)}-\m\a_\rr{\bf 1}\right)$.

\section*{Appendix B: Prefactor}
\renewcommand{\theequation}{B.\arabic{equation}}
\setcounter{equation}{0}

In this Appendix we show that the fermionic constituents of
the prefactor, ${\cal Y}_{\Theta}$,
${\cal W}_{\Lambda}$ and ${\cal W}_{\Theta}$, commute with the kinematical
constraints \eq{feq378}. 
The fermionic constituents of the prefactor have to satisfy 
(the bosonic commutators are trivially zero)
\bea
&&\Big\{ \sum_{\rr=1}^3 \sum_{n \in {\rm\bf Z}}  \a_{\rr} X^{(\rr)}_{mn}
\th_{n(\rr)} \, , {\cal Y}_{\Theta} \Big\} =0, \qquad
\Big\{ \sum_{\rr=1}^3 \sum_{n \in {\rm\bf Z} } X^{(\rr)}_{mn}
\l_{n(\rr)} \, , {\cal Y}_{\Theta} \Big\} =0~, \label{c1} \\
&&\Big\{ \sum_{\rr=1}^3 \sum_{n \in {\rm\bf Z} } \a_\rr X^{(\rr)}_{mn}
\th_{n(\rr)} \, , {\cal W}_{\Lambda (\Theta)} \Big\} =0, \,\,\,
\Big\{ \sum_{\rr=1}^3 \sum_{n \in {\rm\bf Z} } X^{(\rr)}_{mn}
\l_{n(\rr)} \, , {\cal W}_{\Lambda (\Theta)} \Big\} =0, \label{c2} 
\eea
where we expanded the fermionic constraints in oscillation modes.
As usual \cite{Pankiewicz:2002gs, Pankiewicz:2002tg, Chu:2002wj}
it is convenient to split the above equations for $m>0$, $m<0$ and $m=0$.

Consider first the anticommutators of ${\cal W}_{\Lambda}$.
Since ${\cal W}_{\Lambda}$ only contains $b^{\dagger}_{-n}$ oscillators
the non-trivial anticommutators are those with the constraints containing $b_{-n}$. 
By using the mode expansions of $\theta$ and $\lambda$, one obtains for $m>0$
\beq
\label{Wl1}
\Big\{ \sum_{\rr=1}^3 \sum_{n=1}^{\infty}  X^{(\rr)}_{mn}
\l_{n(\rr)} \, , {\cal W}_{\Lambda} \Big\} =
- \frac{1}{2\sqrt{\a'}} \frac{\a}{1-4\mu \a K} \sum_{\rr=1}^3 
\frac{1}{\a_{\rr}}(C^{1/2} A^{\rr} C^{3/2} N^{\rr})_{m} 
= 0
\eeq
because of  Eq.~(5.12) of \cite{Pankiewicz:2002gs}. For $m<0$ 
\bea 
\label{Wl2}
\Big\{ \sum_{\rr=1}^3 \sum_{n=1}^{\infty} \a_\rr X^{(\rr)}_{-m-n}
\th_{-n(\rr)} \, , {\cal W}_{\Lambda} \Big\} &=&  -\frac{i}{\sqrt{\a'}}  
\frac{\a}{1-4\mu \a K} \sum_{\rr=1}^3  C^{1/2}_m 
\Big[\frac{1}{\a_{\rr}^2} (A^{\rr} C^{3/2} C_{(\rr)} N^{\rr})_{m} \nonumber\\
& &+ \frac{\mu}{\a_{\rr}} (A^{\rr} C^{3/2} N^{\rr})_{m}  \Big] = 0
\eea
The two terms in this equation vanish because of Eq.~(5.14) and (5.12)
of \cite{Pankiewicz:2002gs}.

For ${\cal W}_{\Theta}$, the situation is very similar. 
${\cal W}_{\Theta}$ only contains $b^{\dagger}_{n}$ oscillators, so the
non-trivial anticommutators are
\beq
\label{Wt1}
\Big\{ \sum_{\rr=1}^3 \sum_{n=1}^{\infty} \a_\rr X^{(\rr)}_{mn}
\th_{n(\rr)} \, , {\cal W}_{\Theta} \Big\} \,\,\, {\rm for} \,\, m>0~,
\quad
\Big\{ \sum_{\rr=1}^3 \sum_{n=1}^{\infty}  X^{(\rr)}_{-m-n}
\l_{-n(\rr)} \, , {\cal W}_{\Lambda} \Big\} \,\,\, {\rm for} \,\, m<0~,
\eeq
which give respectively Eq.~\eq{Wl1} and \eq{Wl2}.

Finally we have to compute the anticommutator of ${\cal Y}_{\Theta}$. In this case
also the contribution of the zero modes has to be
taken into account.
For $m>0$ there is only one non-trivial anticommutator
\beq
\label{Yt1}
\Big\{ \sum_{\rr=1}^3 \sum_{n=0}^{\infty}  X^{(\rr)}_{mn}
\l_{n(\rr)} \, , {\cal Y}_{\Theta} \Big\} = 
\frac{\a}{2\sqrt{\a'}} C^{1/2}_m \left[ \frac{1}{1-4\mu \a K} \sum_{\rr=1}^3 
(A^{\rr} C^{1/2} U_{(\rr)}^{-1} N^{\rr})_{m} + B_m \right] = 0\,.
\eeq
To prove Eq.~\eq{Yt1}, it is convenient to rewrite $U_{m (\rr)}^{-1}=  
U_{m (\rr)} + 2 \mu \a_\rr C_m^{-1}$. Then by using Eq.~(B.5) of 
\cite{Pankiewicz:2002gs} in the first term and Eq.~(B.9) in the second, it is
easy to see that \eq{Yt1} vanishes.
For $m<0$ the non-trivial anticommutator reads
\beq
\label{Yt2}
\Big\{ \sum_{\rr=1}^3 \sum_{n=1}^{\infty} \a_\rr X^{(\rr)}_{-m-n}
\th_{-n(\rr)} \, , {\cal Y}_{\Theta} \Big\} = \frac{i}{\sqrt{\a'}}
\frac{\a \, \a_3}{1-4\mu \a K}  \sum_{\rr=1}^3 \frac{1}{\a_{\rr}}
(C^{-1/2} A^{\rr} C^{3/2} N^{\rr})_{m} =0,
\eeq
which is zero by Eq.~(B.12) of \cite{Pankiewicz:2002gs}. For $m=0$ the
anticommutators for the zero modes are easily checked to be zero.

\end{document}